\documentclass[prd,aps,preprint,noshowpacs]{revtex4}
\usepackage{mathptmx}
\usepackage[english]{babel}
\usepackage{amssymb}
\usepackage{amsfonts}
\usepackage{amsmath}
\usepackage{eucal}
\usepackage{graphicx}
\usepackage{epsfig}
\def\erf#1{(\ref{#1})} % For references to formulas
\newcommand{\cA}{{\cal A}}  \newcommand{\cB}{{\cal B}}

  \newcommand{\bbN}{{\mathbb N}}

  \newcommand{\bq}{{\mathbf q}}
\newcommand{\bx}{{\mathbf x}}

\def\wn{{\mathfrak{w}}}  \def\qn{{\mathfrak{q}}}

\newcommand{\be}{\begin{equation}} \newcommand{\ee}{\end{equation}}
\newcommand{\bea}{\begin{eqnarray}} \newcommand{\eea}{\end{eqnarray}}
\newcommand{\beann}{\begin{eqnarray*}}  \newcommand{\eeann}{\end{eqnarray*}}
\newcommand{\bfig}{\begin{figure}} \newcommand{\efig}{\end{figure}}
\newcommand{\ba}{\begin{array}} \newcommand{\ea}{\end{array}}
\newcommand{\bcen}{\begin{center}} \newcommand{\ecen}{\end{center}}
\newcommand{\btab}{\begin{tabular}} \newcommand{\etab}{\end{tabular}}

         \def\SU{\operatorname{SU}}

\renewcommand{\Re}{\mathop{\rm Re}}   \renewcommand{\Im}{\mathop{\rm Im}}
\newcommand{\vev}[1]{\left\langle{#1}\right\rangle}

\newcommand{\dd}{{\rm d}}
\newcommand{\e}{{\rm e}}

  \def\Nfour{{\cal N}\!=\!4}

\newtheorem{Proposition}{Proposition}[section]

\newtheorem{Theorem}{Theorem}[section]
\newtheorem{Lemma}{Lemma}[section]
\newtheorem{Corrolary}{Corrolary}[section]

\newcommand{\bp}{\begin{Proposition}}	\newcommand{\ep}{\end{Proposition}}
\newcommand{\bt}{\begin{Theorem}}	\newcommand{\et}{\end{Theorem}}
\newcommand{\bl}{\begin{Lemma}}		\newcommand{\el}{\end{Lemma}}
\newcommand{\bc}{\begin{Corrolary}}	\newcommand{\ec}{\end{Corrolary}}
\begin{document}

%%%%%%%%%%%%%%%%%%%%%%%%%%%%%%%%%%%%%%
%%%%%%%%%%%%% TITLEPAGE %%%%%%%%%%%%%%
%%%%%%%%%%%%%%%%%%%%%%%%%%%%%%%%%%%%%%
\title{Residues of Correlators in the Strongly Coupled $\Nfour$ Plasma}

\author{Irene Amado}\email{Irene.Amado@uam.es}
\author{Karl Landsteiner}\email{Karl.Landsteiner@uam.es}
\author{Sergio Montero}\email{Sergio.Montero@uam.es}
\affiliation{Instituto de F\'{\i}sica Te\'orica UAM/CSIC, Facultad de Ciencias C-XVI,\\
Universidad Aut\'onoma de Madrid, Cantoblanco, Madrid 28049, Spain}
\author{Carlos Hoyos}\email{C.H.Badajoz@swansea.ac.uk}
\affiliation{Department of Physics, Swansea University, Swansea, SA2 8PP, UK}

\begin{abstract}
Quasinormal modes of asymptotically AdS black holes can be interpreted as poles of retarded correlators in the dual gauge theory. To determine the response of the system to small external perturbations it is not enough to know the location of the poles: one also needs to know the residues. We compute them for R-charge currents and find that they are complex except for the hydrodynamic mode, whose residue is purely imaginary.  For different quasinormal modes the residue grows with momentum $\qn$, whereas for the hydrodynamic mode it behaves as a damped oscillation with distinct zeroes at finite $\qn$. Similar to collective excitations at weak coupling the hydrodynamic mode decouples at short wavelengths. Knowledge of the residues allows as well to define the time scale $\tau_{\,\rm H}$ from when on the system enters the hydrodynamic regime, restricting the validity of hydrodynamic simulations to times $t>\tau_{\,\rm H}$.
\end{abstract}

\preprint{IFT-UAM/CSIC-07-53}
\maketitle
%
%%%%%%%%%%%%%%%%%%%%%%%%%%%%%%%%%%%%%%
%%%%%%%%%%%% INTRODUCTION %%%%%%%%%%%%
%%%%%%%%%%%%%%%%%%%%%%%%%%%%%%%%%%%%%%
\section{\label{sec:intro}Introduction}
In recent years a new paradigm concerning the high temperature behaviour of QCD has been established: the strongy coupled Quark-Gluon Plasma (sQGP). Experimental results from heavy ion collisions at RHIC indicate that QCD at temperatures around $2T_c$ is strongly interacting, in spite of being in a deconfined phase, and thus rendering perturbative computations not suitable for describing it. While static properties of strongly coupled gauge theories at finite temperature can be readily analyzed on the lattice, the study of out-of-equilibrium phenomena faces considerable difficulties. In the last years the AdS/CFT correspondence \cite{firstadscft} has emerged as a useful tool to understand analytically the dynamics of non-Abelian gauge theories in the strongly coupled plasma phase.

Perturbative computations in gauge theories show that, at finite temperature, zero momentum correlators of gauge-invariant operators have an inifinite set of discrete poles in the complex frequency plane \cite{Hartnoll:2005ju}. These are responsible of the exponential decay of correlators with time, so describe the dissipation of small perturbations in the plasma. According to the AdS/CFT conjecture, the asymptotically AdS black hole is dual to the plasma phase of the  strongly coupled $\Nfour$ gauge theory \cite{Witten:1998zw}. Correlation functions in the thermal gauge theory can be computed using classical solutions of fields living in the gravity dual \cite{Son:2002sd}. If a field is excited in the presence of a black hole, the energy of the fluctuation will be lost inside the horizon and eventually the final state will be a larger black hole with no fluctuations. This process is described by the quasinormal spectrum, that was first computed in \cite{Horowitz:1999jd} for
  black holes in asymptotically Anti de Sitter (AdS) spacetimes. Quasinormal modes (at fixed momentum) exist only for a discrete set of complex frequencies that can be identified with the poles of the retarded Green's functions in the dual field theory \cite{Grpoles,Nunez:2003eq} . Therefore, quasinormal modes describe dissipation processes in the plasma. 

Hydrodynamic modes like diffusion, shear or sound modes are also described by quasinormal modes in the gravity dual. One of the most interesting results has been the derivation of a universal bound for the shear viscosity to entropy ratio $\frac{\eta}{s} \geq \frac{\hbar}{4\pi k_B}$ \cite{viscosity}. It has also been argued that this value is relevant to the description of heavy-ion phenomenology at RHIC \cite{Muller:2006ee}. One way to derive it is to compute the lowest quasinormal mode in the retarded two-point correlator of the stress tensor. It is an example of a hydrodynamic mode; a special case of quasinormal mode whose frequency vanishes in the zero momentum limit \cite{hydrodynamics}. Another interesting application used the spectral function of R-charge current correlators to derive photon and dilepton production rates in the strongly coupled plasma \cite{CaronHuot:2006te}.

It is helpful to remember the situation at weak coupling for correlators of field components: the poles of the retarded Green's functions correspond to quasiparticle excitations \footnote{At weak couping one also finds branch cuts corresponding to e.g. Landau damping.}. A hard scale of order $T$ can be distinguished from a soft scale of order $gT$. At the hard scale the relevant excitations are the hard partons, i.e. the quarks and gluons. At the soft scale however there are quasiparticles corresponding to the dressed partons and collective excitations, e.g. the longitudinal plasmon mode. Both types of poles can be distinguished by the behaviour of their residues at short wavelengths: the residues of the particle poles scale like $q^{-1}$ with the momentum $q$ whereas the collective excitations show exponentially decaying residues of the form $\exp( -\alpha q^2/g^2T^2 )$ with a mode dependent constant $\alpha$ \cite{leBellac}. To gain a better understanding of the quasinormal
  frequencies that appear in the holographic model of the strongly coupled plasma it is of utmost importance to study the behaviour of their residues. In particular we will concentrate on the transverse and longitudinal R-current correlators. The corresponding quasinormal frequencies and spectral functions have been studied before in \cite{Kovtun:2005ev} and \cite{spectralf,Myers:2007we}, respectively.

Knowledge of the residues at these frequencies is necessary to compute the response of the system to an external perturbation. It also allows to define the time $\tau_{\,\rm H}$ from which on a hydrodynamic approximation is valid. Since hydrodynamic simulations of the evolution of the Quark-Gluon plasma are at the heart of the new sQGP paradigm, it is of utmost importance to know the hydrodynamic time scale $\tau_{\,\rm H}$, which in turn requires knowledge of the residues of the quasinormal modes.

%%%%%%%%%%%%%%%%%%%%%%%%%%%%%%%%%%%%%%
%%%%%%%%%% LINEAR RESPONSE %%%%%%%%%%%
%%%%%%%%%%%%%%%%%%%%%%%%%%%%%%%%%%%%%%
\section{\label{sec:linearesp}Linear Response}
The response to an external perturbation represented by the source $j(t,\bx)$ is given by
\begin{equation}
\vev{\Phi(t,\bx)} =-\int \dd\tau\,\dd^3\xi \,G_{\rm R}(t-\tau, \bx-\xi) \,j(\tau,\xi) ~,
\end{equation}
where $G_{\rm R}$ is the retarded two-point function. This expression is valid when the energy of the perturbation is negligible compared to the total energy of the system. If we choose a perturbation that is localized in time, i.e. we strike the medium once at time $t=0$ and with a periodic profile such that $j(t,\bx) = \delta(t) \cos(\bq\,\bx)$, we find
\begin{equation}\label{eq:linear_response}
\vev{\Phi(t,\bx)} = i\theta(t)\cos(\bq\,\bx) \sum_{n} R_n(\bq) \,\e^{ -i\Omega_n(\bq) t -\Gamma_n(\bq) t} ~,
\end{equation}
where we have assumed that the only singularities of the retarded Green function on the complex frequency plane are poles at the values $\omega_n = \Omega_n -i\,\Gamma_n$ and $R_n$ are the residues. Stability demands that all poles lie in the lower half plane. The contribution of a particular pole is proportional to the residue; hence their importance.

There is a restriction on the location of the poles and on the values of the residues. This restriction can be derived either from the condition that the response \erf{eq:linear_response} has to be real or from symmetry under time reversal $\rho(\omega,\bq)=-\rho(-\omega,\bq)$ where $\rho(\omega,\bq) =-2\Im\widetilde{G}_{\rm R}(\omega,\bq)$. Both conditions lead to the result that the poles come in pairs  and that the residues at a pair are related according to
\begin{equation}\label{eq:reality}
\widetilde{\omega}_n = -\omega_n^* ~,\quad \widetilde{R}_n = - R_n^* ~.
\end{equation}
There can also be unpaired poles lying on the imaginary axes and in this case their residues have to be purely imaginary. 

We are able to compute the location of the poles and the value of the residues using the gravity dual. They are determined by the values of the frequency where quasinormal modes exist, and by the shape of these classical solutions. In our numerical calculations we have checked that the stability and time reversal relations indeed hold for the quasinormal modes. Analytic proofs can be found in \cite{Amado:2007pv}.

Of particular interest are the hydrodynamic modes, whose frequencies
vanish as the momentum goes to zero. At sufficiently small momentum the hydrodynamic mode has the smallest imaginary part of all poles and therefore dominates the long time behaviour. Knowledge of the residues for the quasinormal modes allows us to define a hydrodynamic time scale $\tau_{\,\rm H}$, from which on the hydrodynamic mode dominates the response\footnote{In kinetic theory at weak coupling the validity of the hydrodynamic approximation can be defined through the ratio of mean free time to the time scale of the process under consideration, often referred to as Knudsen number. Hydrodynamics is valid for Knudsen numbers much smaller than one.}. From \erf{eq:linear_response} we find
\begin{equation}
\tau_{\,\rm H} = \frac{\log|R_{\rm H}|-\log|R_1|}{\Gamma_{\rm H}-\Gamma_1} ~.
\end{equation}
Only for times larger than $\tau_{\,\rm H}$ we can expect hydrodynamics to be a good approximation. For shorter times more and more higher poles will contribute. 

%%%%%%%%%%%%%%%%%%%%%%%%%%%%%%%%%%%%%%
%%%%%% EXACTLY SOLVABLE EXAMPLE %%%%%%
%%%%%%%%%%%%%%%%%%%%%%%%%%%%%%%%%%%%%%
\section{\label{sec:exactlysolv}An exactly solvable example}
In general the quasinormal modes of the five dimensional AdS black hole can be determined only numerically. There are however some special cases where the wave equations simplify and can be solved analytically. Two such cases are the wave equations for the gauge invariant variables corresponding to longitudinal and transverse vector field perturbations $E_L = k A_0 + \omega\,\bq\cdot\mathbf{A}/q$, $E_T = \omega A_T$. The AdS black hole background is given by
\begin{equation}
\dd s^2 = \frac{r^2}{L^2} \left(- f(r) \dd t^2 + \dd\bx^2 \right)  + \frac{L^2}{r^2}\frac{\dd r^2}{f(r)} ~,
\end{equation}
with $f(r)= (1-r_0^4/r^4)$. The Hawking temperature $T_H = r_0/\pi L^2$ is the temperature $T$ of the dual field theory. In the following we will use the coordinate $x=1-r_0^2/r^2$ such that the horizon sits at $x=0$ and the boundary at $x=1$.
The equations of motion for these gauge invariant combinations of the vector field perturbations with frequency $\omega$ and momentum $q$ are \cite{Kovtun:2005ev}
\begin{subequations}\label{eq:vectorfields}
\begin{eqnarray}
E_T'' + \frac{f'}{f} \,E_T' + \frac{\wn^2- f \qn^2}{(1-x) f^2} \,E_T &=&0 ~, \\
E_L'' + \frac{\wn^2 f'}{f(\wn^2-f \qn^2)} \,E_L' + \frac{\wn^2 - f \qn^2}{(1-x) f^2} \,E_L &=& 0 ~,
\end{eqnarray}
\end{subequations}
where we have introduced the dimensionless frequency and momentum $2\pi T (\wn,\qn) :=(\omega,\bq)$.
According to the holographic dictionary the vector field in AdS acts as a source for the
R-charge currents $J^a_\mu$ of the $\SU(4)$ R-symmetry in the dual gauge theory. 
Denoting the retarded correlator of
two currents as $G_{\mu \nu}$ one can write for the non-vanishing components \cite{Kovtun:2005ev}
\begin{eqnarray}\label{eq:Greensfunctions}
\widetilde{G}_{TT} &=& \Pi^T(\omega, \bq) ~,\hspace*{1em} \widetilde{G}_{tt} = \frac{\bq^2}{\omega^2-\bq^2} \Pi^L(\omega,\bq) ~, \\
\widetilde{G}_{tL} &=& \frac{-q \omega}{\omega^2-\bq^2} \Pi^L(\omega,\bq) ~,\hspace*{1em}  \widetilde{G}_{LL} = \frac{\omega^2}{\omega^2-\bq^2} \Pi^L(\omega,\bq) ~. \nonumber
\end{eqnarray}
The correlators are therefore defined by the transverse and longitudinal polarization tensors $\Pi^{T,L}(\omega, \bq)$.

At zero momentum $\bq=0$ longitudinal and transverse components become indistinguishable from each other and obey the same wave equation. Moreover at $\qn =0$, equations \erf{eq:vectorfields} can be solved exactly and the polarization tensor was found to be \cite{Myers:2007we}
\begin{equation}\label{eq:pi}
\frac{\Pi(\wn)}{N^2 T^2/8} = i \wn  + \wn^2 \left[\psi\left( \frac{(1-i)}{2}\wn\right) + \psi\left( \frac{-(1+i)}{2}\wn\right) \right] ~,
\end{equation}
where $\psi$ is the digamma function and $N$ the number of colors. The locations of the poles and their residues are
\begin{equation}\label{eq:qnms_exact}
\omega_n^\pm = 2 \pi T n( \pm 1 - i) ~,\quad R_n^\pm =  \frac{\pi}{2} N^2 T^3 n^2 (\mp 1 - i) ~.
\end{equation}
with $n\in \bbN $. The residues fulfil the reality conditions \erf{eq:reality}. At first glance it looks strange that the residues grow quadratically with the mode number. If we want to write the retarded Green's function as a sum over the poles this leads to a divergent expression. However, we have to remember that the poles determine the Green's function only up to an analytic part and that this analytic part can be infinite. In order to obtain a well defined pole representation one has therefore to substract a conveniently chosen analytic part. This is made explicit in the well-known pole representation of the digamma function
\begin{equation}\label{eq:digammaexp}
\psi(x) = -\gamma_{\ \rm E} -\sum_{n=1}^\infty \left( \frac{1}{x-1+n} -\frac{1}{n} \right) ~,
\end{equation}
where $\gamma_{\ \rm E}$ is the Euler--Mascheroni number.
Using this we can find a pole representation of the polarization tensor as
\begin{equation}\label{eq:piexp}
\Pi(\wn) = \frac{N^2 T^2}{8} \Bigg\{ -i \wn-2 \gamma_{\ \rm E} \,\wn^2  +\wn^3 \sum_{n=1}^\infty \frac{1}{n} \left( \frac{1}{\wn-n(1-i)} +\frac{1}{\wn+n(1+i)} \right) \Bigg\} ~.
\end{equation}
The linear term in $\wn$ can be understood as the zero momentum limit of the diffusion pole. This suggests that the retarded Green's function at non-zero momentum would be a meromorphic function of frequency and momentum and admit an expansion
\begin{equation}
\Pi(\wn, \qn)= \sum_n \frac{R_n(\wn, \qn)}{\wn-\wn_n(\qn)} ~,
\end{equation}
where $R_n(\wn,\qn )$ are analytic functions. We will do a numerical evaluation of the residues at the pole $R_n(\wn_n(\qn),\qn )$. This implies that our method does not disentangle the frequency from the momentum dependence.

A rather interesting point is to ask what is the spectral line produced by a single isolated quasinormal mode. Due to the fact that the residue is complex the line spectrum of an isolated quasinormal mode does not have the shape of a Lorentzian function. Writing $Y=\Im(R)/\Re(R)$, we find
\begin{equation}
\rho_{\rm QNM}(\wn) \propto \,\frac{\Gamma -Y(\wn-\Omega)}{(\wn-\Omega)^2 + \Gamma^2} ~.
\end{equation}
For non-vanishing imaginary part of the residue the peak is higher than the one of a Lorentz curve with $Y=0$ and it is shifted to lower frequencies if $Y>0$ or to higher frequencies if $Y<0$. In the case at hand where the $\Gamma /\Omega \approx 1$ the quasiparticle approximation is not valid and the line spectrum of an isolated quasinormal mode does not approximate the full spectral function in any range of the frequency. In cases where $\Gamma/\Omega \ll 1$ and $Y$ different from zero it would show up as an asymmetry in the form of the observed resonance in the spectral function and a correct interpretation is possible only if the residue is known~\footnote{Narrow peaks in holographic spectral functions have been recently observed in \cite{Myers:2007we,Erdmenger:2007ja}.}.

The expression found as a sum over poles and the exact one differ only by some contact terms, coming from the analytic pieces we have neglected. Neglecting the contact terms, the Green's function diverges at $t=0$ as $t^{-3}$, coinciding with the zero-temperature result. This shows that the sum over poles is enough to recover most of the dynamical information and only the contact terms require a more detailed analysis. In principle they should be related to the Schwinger terms of the $T=0$ theory.

%%%%%%%%%%%%%%%%%%%%%%%%%%%%%%%%%%%%%%
%%%%%%%% NUMERICAL ALGORITHM %%%%%%%%%
%%%%%%%%%%%%%%%%%%%%%%%%%%%%%%%%%%%%%%
\section{\label{sec:numerical}Numerical Computation}
\begin{figure}[!htbp]
\includegraphics[scale=1.2]{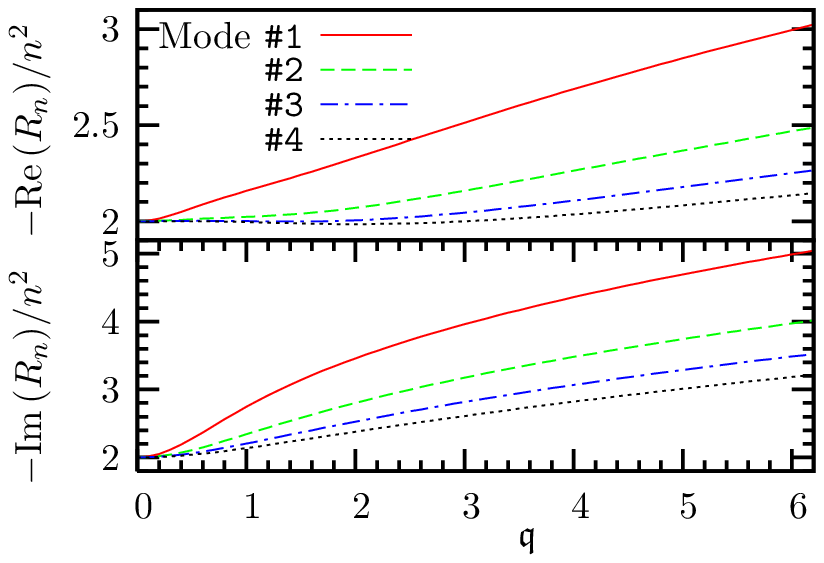}\\[2ex]
\includegraphics[scale=1.2]{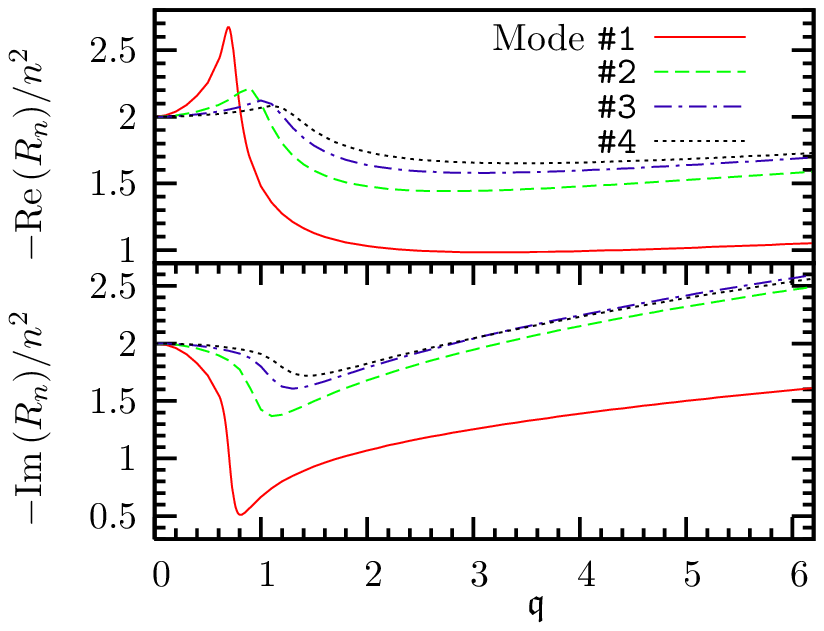}
\caption{\label{fig:ResFigs}(Top) Real and imaginary parts of the residues for the first four quasinormal modes in the transverse component $E_T$ (bottom) Idem for longitudinal component $E_L$.  The $n^2$ scaling is necessary to recover the asymptotic behaviour of the spectral function at large frequencies. Close to the crossing with the diffusion mode $\qn\sim 1$, the residues of the longitudinal component present peaks. The residues grow with momentum, this is also reflected in the growth of the spectral function.}
\end{figure}

\begin{figure}[!htbp]
\includegraphics[scale=1.2]{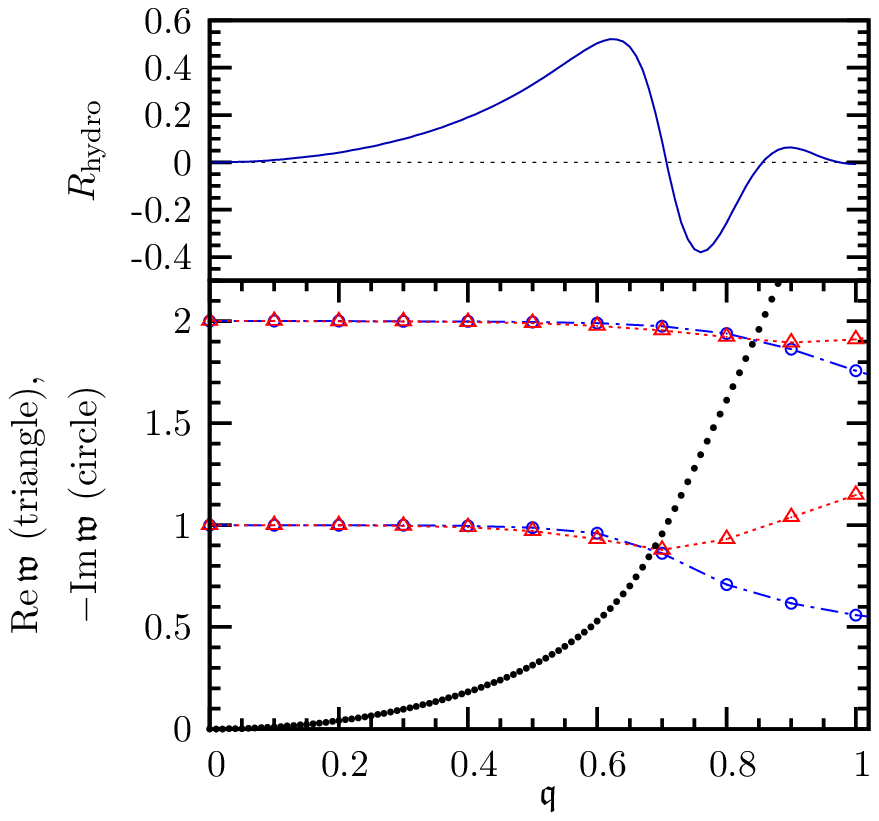}
\caption{\label{fig:Crossing} In the lower figure, we show the real and imaginary parts of the quasinormal frequencies in the longitudinal channel and the value of the frequency for the diffusion mode. The quasinormal frequencies remain fairly stable as momentum increases, until there is a 'crossing' with the diffusion mode (it reaches a special value $\wn=-in$). Then, there is a qualitative change in the behaviour of the quasinormal frequency that starts approaching the real axis. This behaviour can also be oberved in previous computations of quasinormal frequencies \cite{Nunez:2003eq}. The residue of the diffusion mode, shown in the upper figure, behaves according to hydrodynamics $\sim q^2$ for small momentum. For larger momentum, it shows an oscillatory and decaying behaviour. The zeroes of the residue coincide with the 'crossing' values. Quasinormal frequencies in the transverse channel approach smoothly the real axis as momentum increases.}
\end{figure}

According to the holographic dictionary the retarded Green's function can be computed as the
ratio of the connection coefficients that relate the local solution at the horizon with
ingoing boundary conditions to the non-normalizable $(\cA)$ and normalizable $(\cB)$ solutions at the boundary. Defining $(\alpha):=(T,L)$ as the two components
\begin{equation}
E_{(\alpha)}^{\rm h}(x) = \cA_{(\alpha)} \,E_{(\alpha)}^1(x) + \cB_{(\alpha)} \,E_{(\alpha)}^2(x) ~.
\end{equation}
The retarded polarization tensors are given then by \cite{Kovtun:2005ev}
\begin{equation}
\Pi^{(\alpha)} = -\frac{N^2 T^2}{8} \,\frac{\cB_{(\alpha)}}{\cA_{(\alpha)}} ~.
\end{equation}
The quasinormal modes are normalizable solutions where the connection coefficient $\cA_{(\alpha)}=0$.
The polarization tensor follows from demanding that the solution is smooth at a matching point in the interior of the interval $(0,1)$,
\begin{equation}\label{eq:connection_coeffs}
\frac{\cB_{(\alpha)}}{\cA_{(\alpha)}} = \frac{ E_{(\alpha)}^{\rm h} (E_{(\alpha)}^1)' - (E_{(\alpha)}^{\rm h})' E_{(\alpha)}^1}{ E_{(\alpha)}^2 (E_{(\alpha)}^{\rm h})' - (E_{(\alpha)}^2)' E_{(\alpha)}^{\rm h} } ~.
\end{equation}
We have computed the Frobenius series up to order 50. Matching the series expansions, we see 
that the ratio (\ref{eq:connection_coeffs}) remains constant for a fair interval in the radial coordinate. We have chosen $x=0.53$ to evaluate the ratio and have checked that the spectral function agrees with previous numerical (for non-zero momentum) and exact (for zero momentum) results \cite{spectralf, Myers:2007we}.
The residue can be computed as
\begin{equation}
R_n^{(\alpha)} = \frac{\pi}{4} N^2 T^3\left[ \left. \frac{\partial}{\partial \wn} \left(\frac{\cA_{(\alpha)}}{\cB_{(\alpha)}}\right)\right|_{\wn=\wn_n} \right]^{-1} ~.
\end{equation}
Our results for the residues (normalized by $[(\pi N^2 T^3 n^2)/4]^{-1}$) of transverse and longitudinal vector fluctuations are plotted in Figure \ref{fig:ResFigs}. The longitudinal fluctuations show interesting behaviour related to the diffusion mode. The peaks and dips in Figure \ref{fig:ResFigs} appear roughly at the locations where the hydrodynamic mode\,\footnote{We continue calling it like this even outside the regime where the actual hydrodynamic approximation is valid.} crosses the imaginary part of the quasinormal frequency. The hydrodynamic mode behaves in a completely different way. The diffusion pole quickly moves towards negative imaginary frequencies, while the residue first grows according to hydrodynamics and later goes over into a damped oscillation. Numerically the zeroes coincide with $\wn=-i n$. On the other hand, the location of the other quasinormal modes remains fairly constant until the diffusion pole reaches an integer value that coincides with the imaginar
 y part of the quasinormal pole at zero momentum. After this point, the quasinormal pole starts approaching the real axis. This first happens at $\qn^2=1/2$, when the diffusion pole is at $\wn=-i$. For larger momentum, the long time behaviour of the fluctuations will be dominated by the first quasinormal mode and not by the hydrodynamic mode. We also find some interesting analytic structure related to the zeroes of the residue of the hydrodynamic pole (see Figure \ref{fig:Crossing}). When $\wn=-i n$, the exponents of ingoing and outgoing solutions at the horizon differ by an integer. The ingoing solution should then develop a logarithmic term since it has the lower exponent. However, for very special values of the parameters, i.e. the momentum $\qn$, it can happen that the coefficient of the logarithm vanishes. We find numerically that the zeroes of the residues of the hydrodynamic mode coincide with these special values at $(\wn, \qn^2) = (-i,1/2)$, $(-2i,\sqrt{3}-1)$, $(-3 
 i,\sqrt{6}-3/2)$ and $(-4 i, 1.4436)$.

%%%%%%%%%%%%%%%%%%%%%%%%%%%%%%%%%%%%%%
%%%%%%%%%%%%%%% DISCUSSION %%%%%%%%%%%%%%
%%%%%%%%%%%%%%%%%%%%%%%%%%%%%%%%%%%%%%
\section{\label{sec:results}Discussion}
The analytic structure found for the R-charge correlator at finite temperature is very interesting. At zero momentum, an infinite set of evenly spaced poles on the complex frequency plane is a generic feature of correlators. At weak coupling, they lie on the imaginary axis \cite{Hartnoll:2005ju}, but according to the AdS/CFT computation they move away at strong coupling. The right UV behavior, that is analytic at zero temperature, is recovered after summing over all the poles. The value of the residues is crucial, and especially the vanishing of the hydrodynamic mode at zero momentum. When we consider momentum dependence, the analytic form of the diffusion residue in the hydrodynamic approximation $\sim q^2$ will induce a singularity in the correlator at short times, on top of the usual UV singularity. This clearly cannot be the right answer, and we can understand this as a limit in the validity of the approximation. The damped and oscillatory behavior of the hydrodynamic residue can cure this problem, making the contribution from the diffusion mode smooth at short times. The hydrodynamic mode effectively
decouples for momenta $q > 1$ and in this respect behaves as the collective modes present at weak coupling. In this sense, our results go beyond the hydrodynamic approximation. In principle, we expect that other hydrodynamic modes appearing in the stress-tensor two-point functions, the shear and the sound mode, to have a similar collective mode behavior.

We observe that as the momentum increases, diffusion becomes less important and other collective excitations of longitudinal modes describe charge density fluctuations. In contrast with the dressed partons of the weak coupling regime, they do not decouple at high momentum. The behaviour is also different from the poles found for gauge invariant operators. At weak coupling, these poles open up in branch cuts at fixed positions in the imaginary axis, while the holographic computation predicts that at infinite coupling the only singularities are poles that come closer to the real axis. Eventually a new peak appears in the spectral function, located close to $\wn=\qn$. This peak persists at higher momentum, and can be interpreted as a quasiparticle excitation of charge density fluctuations. This shows a change of behaviour of the system as we increase the momentum, from diffusive to reactive. Transverse fluctuations are comparatively featureless, there are no quasiparticle excitations appearing; this reflects the fact that there are no propagating modes in the zero temperature conformal theory.

An application of our results is the calculation of the hydrodynamic time scale $\tau_{\,\rm H}$. Using our numerical results we find that in units of $(2\pi T)^{-1}$ the minimal time scale is $\tau_{\,\rm H}=3.7-3.2$  in a rank $\qn=0.3-0.48$, rapidly growing for higher values of the momentum. For lower values it also grows, but this is due to the fact that the charge distribution is already quite uniform, so we can take this value as the onset of diffusion. In fact the dispersion relation starts to deviate from the hydrodynamic approximation for $q\approx 0.45$ which corresponds to Compton wavelengths of $~1.2$ fm, the size of a proton. As a model for sQGP in RHIC physics, let us pick $T\simeq 2 T_c\simeq 350$ MeV. The hydrodynamic time scale is then $\tau_{\,\rm H} \approx 0.3$ fm/c. This is a remakably short time. In fact it indicates that the hydrodynamic approximation is valid from very short times on. Notice that at RHIC the thermalization time is $\tau_\mathrm{therm} \approx 0.6$ fm/c and the hydrodynamic approximation is therefore valid already for $t\lesssim 1$ fm/c \cite{Muller:2006ee}. We expect that the values for shear or sound modes will be slightly different, but it is reassuring to find the right orders of magnitude even for $\Nfour$.

%%%%%%%%%% Acknowledgements %%%%%%%%%%
\begin{acknowledgments}
K.\,L. is supported by a Ram\'on y Cajal contract. S.\,M. is supported by an FPI 01/0728/2004 grant from Comunidad de Madrid. I.\,A. is supported by grant BES-2007-16830.
I.\,A., K.\,L. and S.\,M. are supported in part by the Plan Nacional de Altas Energ\'{\i}as FPA-2006-05485, FPA-2006-05423 and EC Commission under grant MRTN-CT-2004-005104. S.\,M. wants to thank G. S\'anchez for her support. We would like to thank G. Aarts, D. Kaplan, P. Kumar, E. L\'opez, A. Rebhan and A. Vuorinen for useful discussions.
\end{acknowledgments}

%%%%%%%%%%%%%%%%%%%%%%%%%%%%%%%%%%%%%%
%%%%%%%%%% THE BIBLIOGRAPHY %%%%%%%%%%
%%%%%%%%%%%%%%%%%%%%%%%%%%%%%%%%%%%%%%

\end{document}